# Giant magnetic moment enhancement of nickel nanoparticles embedded in multi-walled carbon nanotubes

Guo-meng Zhao [1,2], Pieder Beeli [1], Jun Wang [2], and Yang Ren[3]

[1] Department of Physics and Astronomy, California State University, Los Angeles, CA 90032, USA

[2] Department of Physics, Faculty of Science, Ningbo University, Ningbo, P. R. China

[3] X-Ray Science Division, Advance Photon Source, Argonne National Laboratory, Argonne, IL 60439, USA

## Abstract

We report a giant magnetic moment enhancement of ferromagnetic nickel nanoparticles (11 nm) embedded in multi-walled carbon nanotubes. High-energy synchrotron x-ray diffraction experiment and chemical analysis are used to accurately determine the ferromagnetic nickel concentration. Magnetic measurements up to 900 K show an intriguing result—the saturation magnetization of the nickel nanoparticles embedded in the multi-walled carbon nanotubes is about 3.4 times as much as the value expected from the measured nickel concentration while the Curie temperature is the same as that of bulk nickel. The implication of this giant magnetic moment enhancement, whatever its origin, is likely to be far reaching—it opens a new avenue for implementing spin-electronics at the molecular level.

Graphene is a sheet of carbon atoms distributed in a honeycomb lattice and is the building block for graphite, carbon nanotubes, and carbon buckyballs. The massless, relativistic behavior of graphene's charge carriers, known as Dirac fermions, is a result of its unique electronic structure, characterized by conical valence and conduction bands that meet at a single point in momentum space. The integer quantum Hall effect recently observed in graphene is closely related to these Dirac fermions (1,2,3). In addition to the remarkable room temperature quantum Hall effect in graphene (3), there are intriguing reports of high-temperature superconducting behaviors in both graphite (4) and graphite-sulfur composites (5,6). Highly oriented pyrolithic graphite (HOPG) has been shown to display either a partial superconducting or a ferromagnetic-like response to an applied magnetic field even at temperatures above room temperature (4). Another intriguing finding is that the measured saturation magnetizations of some graphite samples from the Canyon Diablo meteorite are significantly (36%-110%) larger than the values expected from the magnetic impurity concentrations inferred from Mössbauer spectra (7). The result appears to provide evidence for extra magnetic moments contributed from graphite. This extra contribution to the moment has been attributed to a large magnetic proximity effect at the interface with magnetite or kamacite inclusion, which induces a ferromagnetic ordering in graphite with a Curie temperature of 570 K (7). However, this important conclusion strongly depends on whether the measurement of the magnetic impurity concentrations in this study is accurate enough.

Here we study magnetic properties of multi-walled carbon nanotubes (MWCNTs) embedded with ferromagnetic Ni nanoparticles. Because Ni nanoparticles sit inside the innermost shells of MWCNTs and electronic inter-shell coupling is negligibly small, we expect that the only shells that can interact with the Ni nanoparticles are the innermost shells. Then the extra magnetic moments induced by the magnetic proximity effect should be much smaller than those found in graphite (7). However, our data show a giant magnetic moment enhancement that is about two orders of magnitude larger than that predicted from the magnetic proximity effect.

We used a purified multi-walled carbon nanotube mat sample (Lot No. TS0636) obtained from SES Research of Houston. The sample was prepared by chemical vapor deposition under catalyzation of nickel nanoparticles. After purification, most nickel particles were removed except for those embedded inside the innermost shells. The morphology of the mat sample can be seen from scanning electron microscopy (SEM) images shown in Fig. 1. The SEM images were taken by a field emission scanning electron microcopy (FE-SEM, Hitachi S-4800) using an accelerating voltage of 3 kV. One can clearly see that the outer diameters of these MWCNTs are in the range of 30-50 nm and centered around 40 nm.

The total metal-based impurity concentrations of the mat sample can be determined from the composition analysis of the residual of the sample, which was obtained by burning off carbon-based materials in air. A Perkin–Elmer Elan-DRCe inductively coupled plasma mass spectrometer (ICP-MS) was used to analyze the composition of the residual. From the weight (1.33%) of the residual and the ICP-MS analysis, we obtain the metal-based magnetic impurity concentrations in weight: Ni = 0.476%, Fe = 0.00907%, and Co = 0.0133%. The other major non-magnetic purity is La, whose concentration is found to be 0.457%.

In order to quantitatively understand the magnetism of the MWCNTs embedded with the nickel-based impurities, it is essential to determine the concentration of the ferromagnetic nickel phase. We can achieve this goal by performing high-energy synchrotron x-ray diffraction (XRD) experiment. Figure 2**A** shows a synchrotron XRD spectrum for a virgin sample. The XRD spectrum was taken on a high-energy synchrotron x-ray beam-line 11-ID-C at the Advanced Photon Source, Argonne National Laboratory, using monochromated radiation with a wavelength of $\lambda = 0.1078$ Å. The XRD pattern was collected with a Mar345 image plate detector, and converted to 1D pattern using the FIT2D program (8). It is apparent that the major peaks in the spectrum correspond to the diffraction peaks of MWCNT (9) and the face-centered cubic (FCC) phase of Ni. In particular, the (002) diffraction peak of MWCNT is seen at $2\theta = 1.815°$ and the Ni (311) peak at $2\theta = 5.815°$. Fig. 2**B** and 2**C** display the expanded views of the MWCNT (002) and Ni (311) peaks, respectively. The solid red line in Fig. 2**B** is the fitted curve by the sum of a Gaussian and a Lorenztian function, which takes into account both domain size broadening and strain broadening (9). The fit with the sum of two Gaussian or two Lorenztian functions is not excellent. The solid red line in Fig. 2**C** is the fitted curve by a Gaussian function that is consistent with particle-size broadening (9). The integrated intensity of the Ni (311) peak is found to be 0.84% of the intensity of the MWCNT (002) peak. Using the calculated standard intensities of the graphite (002) and Ni (311) peaks and assuming that the intensity of MWCNT (002) peak is the same as that of graphite (002), we find that the ferromagnetic FCC nickel concentration is 0.43% (in weight), which is slightly lower than the total Ni concentration (0.476%) inferred from ICP-MS. This implies that the ferromagnetic FCC nickel is the dominant phase while the concentrations of other nonmagnetic nickel-based phases are too small to be seen in the XRD spectrum.

In order to check the reliability of our inferred ferromagnetic nickel concentration based on the Ni (311) peak, we show, in Fig. 2**D**, the expected XRD spectrum of the FCC

Ni with the concentration of 0.43% (green blue line) and the difference spectrum (red line), which is obtained by subtracting the Ni spectrum from the spectrum of TS0636 in Fig. 2**A**. The difference spectrum shows no observable residual of any peaks of the FCC nickel, implying that the inferred Ni concentration is indeed reliable. Furthermore, all the peaks except for some peaks indicated by arrows in the difference spectrum agree with the peaks observed and calculated for pure MWCNTs (9). The extra peaks indicated by the arrows should be associated with La-based impurity phases.

Since magnetic properties of nanoparticles depend strongly on the particle size, it is important to determine the average diameter of the ferromagnetic Ni nanoparticles imbedded in MWCNTs. We can determine the average diameter $d$ from the peak width of the XRD spectrum. The full width at half maximum (HWHM) of the Ni (311) peak is found to be 0.0556° from the Gaussian fit in Fig. 2**C**. Using Scherrer's equation (10): $d = 0.89\lambda/(\beta\cos\theta)$ and with $\beta = 0.0511°$ (after correcting for the instrumental broadening), we calculate $d = 11$ nm.

With the information of the average diameter (11 nm) of the ferromagnetic Ni nanoparticles and the ferromagnetic Ni concentration of 0.43% in our MWCNT sample, we would expect the room-temperature saturation magnetization ($M_s$) to be 0.14 emu/g from the known $M_s = 30$-$32$ emu/g for pure FCC Ni nanoparticles with $d =11$-$12$ nm (11). If we use the upper limit of $M_s = 55$ emu/g for bulk Ni, the expected $M_s$ for our MWCNT sample would be 0.24 emu/g. Fig. 3**A** shows magnetization versus magnetic field for our MWCNT mat sample at 320 K. The magnetization was measured using a Quantum Design vibrating sample magnetometer (VSM). The linear field dependence of the magnetization with a negative slope at $H > 10$ kOe is due to the diamagnetic contribution.

The linear extrapolation to $H = 0$ yields $|\chi_{dia}| = 4.2 \times 10^{-6}$ emu/g and $M_s = 0.47$ emu/g. The measured $M_s$ value is a factor of 3.4 larger than the expected value (0.14 emu/g) from the measured Ni concentration and average diameter. Thus, there is a giant magnetic moment enhancement of the Ni nanoparticles when they are embedded inside the innermost shells of MWCNTs. Using the measured nickel concentration (0.43%), the enhanced saturation magnetization $\Delta M_s$ is calculated to be about 76.7 emu per gram of nickel or 683 emu per cubic centimeter of nickel. For our ferromagnetic iron nanoparticles ($d = 55$ nm) embedded in MWCNTs, $\Delta M_s$ is found to be about 273 emu per gram of iron or 2154 emu per cubic centimeter of iron.

Another intriguing result is that magnetization is not reversible at a field higher than 20 kOe, as clearly seen from an expanded view of the magnetization versus magnetic field (Fig. 3**B**). We have confirmed that this irreversibility is not a measurement artifact, but an intrinsic property of the sample.

If the observed giant magnetic moment enhancement arises from the magnetic proximity effect, then as proposed in ref. (7) there should exist *two* distinctive ferromagnetic-like transitions. Fig. 4**A** shows field-cooled magnetic susceptibility of the MWCNT sample in a field of 100 Oe. We clearly see that there is only one ferromagnetic transition at about 630 K. For comparison we show, in Fig. 4**B**, the field-cooled susceptibility for pure nickel particles ($d = 800$ nm). Within the uncertainty ($\pm 20$ K) in the sample temperature, the Curie temperature of the pure nickel particles is the same as that of the nickel nanoparticles embedded in MWCNTs.

Now we turn to discuss the origin of the giant magnetic moment enhancement of the nickel nanoparticles embedded in MWCNTs. One possibility is that this enhancement is likely to arise from a large magnetic proximity effect (7,12). We consider a simple case where our ferromagnetic nanoparticles have a cylindrical shape with both length and diameter equal to $d$ and the curve surface of the cylinder contacts with the innermost shell of a MWCNT. The curved surface area is equal to $\pi d^2$ and the total number of the contact carbon is $\pi N_C d^2$, where $N_C$ is the number of carbon per unit area and equal to $3.82\times10^{15}/cm^2$ (13). If the induced magnetic moment is $m\mu_B$ (where $\mu_B$ is the Bohr magneton) per contact carbon atom, then the induced saturation magnetization normalized to the volume of the ferromagnetic nanoparticle is $\Delta M_s = 4N_C m\mu_B/d = 1420(m/d)$ emu/cm$^3$ (here $d$ is in units of nm). Using the measured $\Delta M_s = 683$ emu/cm$^3$ and $d = 11$ nm for ferromagnetic nickel nanoparticles, we find that $m = 5.3$, which is a factor of 53 larger than the value (~0.1) calculated using density function theory (12). For ferromagnetic iron nanoparticles with $d = 55$ nm, the measured $\Delta M_s = 2154$ emu/cm$^3$ implies $m = 83.4$. Therefore, the magnetic proximity model seems unlikely to explain such a giant magnetic moment enhancement. A new theoretical model is required to consistently explain both ferromagnetic and superconducting-like hysteresis loops observed in HOPG at 400 K (4) as well as our current intriguing results involving both a giant magnetic moment enhancement and an unusual high-field hysteresis effect shown in Fig. 3**B**.

The implication of this giant magnetic moment enhancement, whatever its origin, is likely to be far-reaching—it may open a new avenue for implementing spin-electronics at the molecular level. Carbon nanotubes have been demonstrated to have important applications in logic circuits (14–16) and non-volatile random access memories (17). The strong interplay between the ferromagnetic nanoparticles and carbon nanotubes can open new perspectives for spin-based high-speed devices, where logic and memory elements are integrated at the molecular level.

We thank M. Du and F. M. Zhou in the Department of Chemistry and Biochemistry at CSULA for the elemental analyses using ICP-MS. We also thank the Palmdale Institute of Technology for the use of the VSM. Use of the Advanced Photon Source was supported by the U.S. Department of Energy, Office of Science, Office of Basic Energy Sciences, under Contract No. DE-AC02-06CH11357.

**Figures**

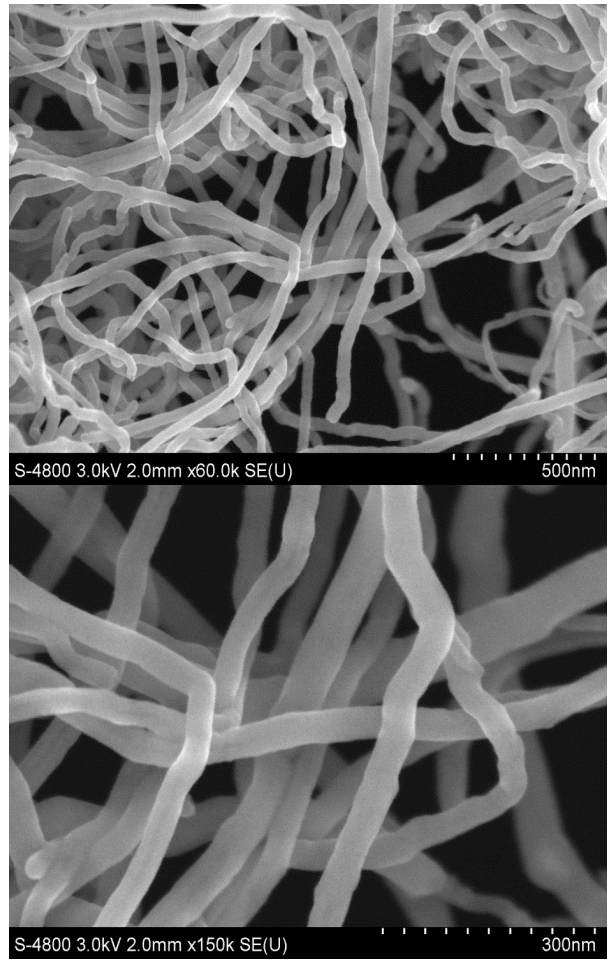

Fig. 1. Scanning electron microscopic (SEM) images of a multi-walled carbon nanotube mat sample. The SEM images were taken by a field emission scanning electron microcopy (FE-SEM, Hitachi S-4800) using an accelerating voltage of 3 kV. The outer diameters of these MWCNTs are in the range of 30-50 nm and centered around 40 nm.

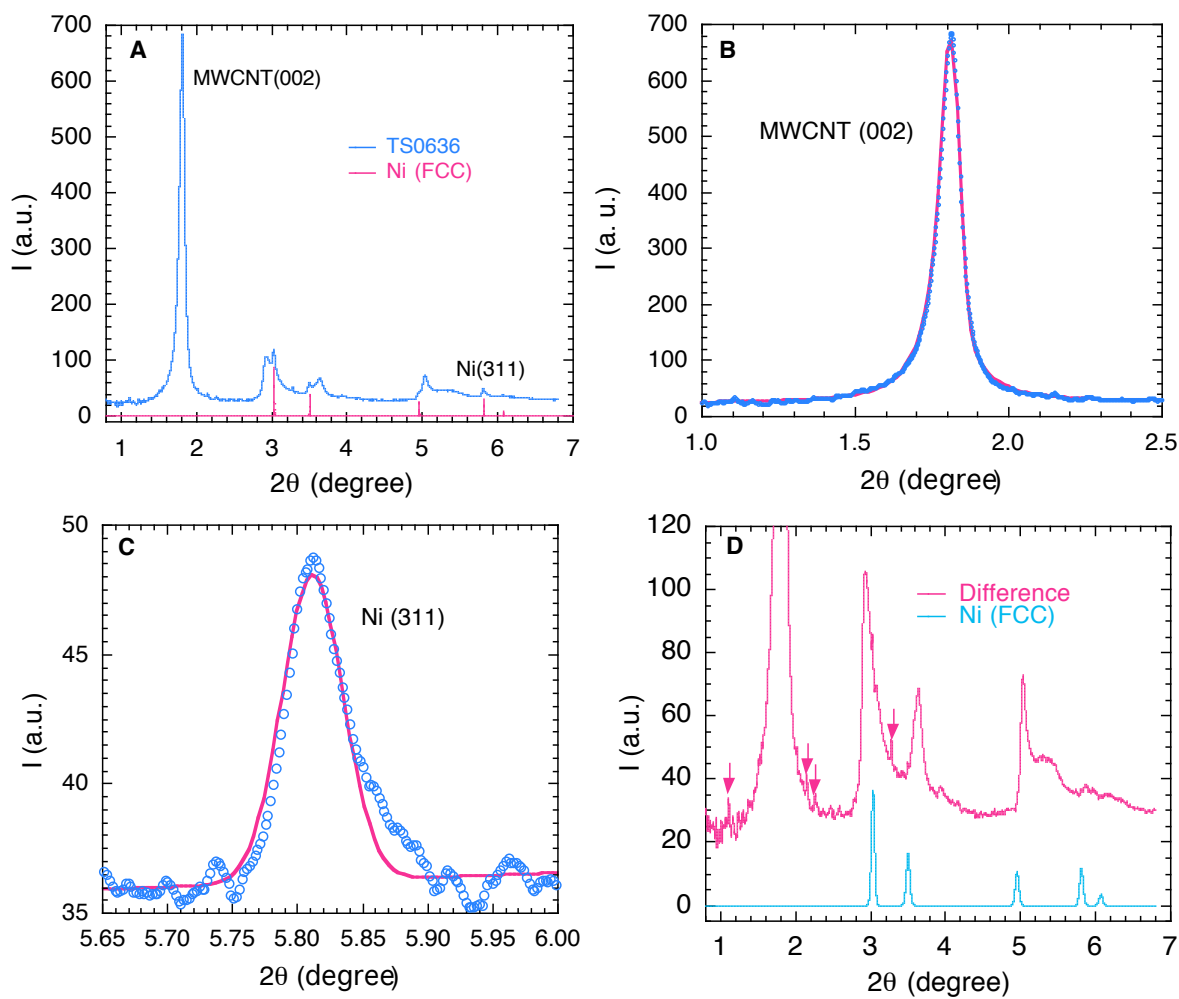

Fig. 2. High-energy synchrotron x-ray diffraction (XRD) spectra. (**A**) XRD spectrum for a virgin mat sample of multi-walled carbon nanotubes (TS0636) and the standard spectrum for face-centered cubic phase of nickel. The wavelength of the high-energy x-ray radiation is 0.1078 Å. (**B**) The expanded view of the MWCNT (002) peak. The solid red line is the fitted curve by the sum of a Gaussian and a Lorenztian function. (**C**) The expanded view of the Ni (311) peak. The solid red line is the fitted curve by a Gaussian function. (**D**) The expected XRD spectrum of the FCC Ni (green blue line) based on the nickel concentration (0.43%) and the difference spectrum (red line), which is obtained by subtracting the Ni spectrum from the spectrum of TS0636 in (**A**). All the peaks except for some peaks indicated by arrows in the difference spectrum agree with the peaks observed and calculated for pure MWCNTs (9).

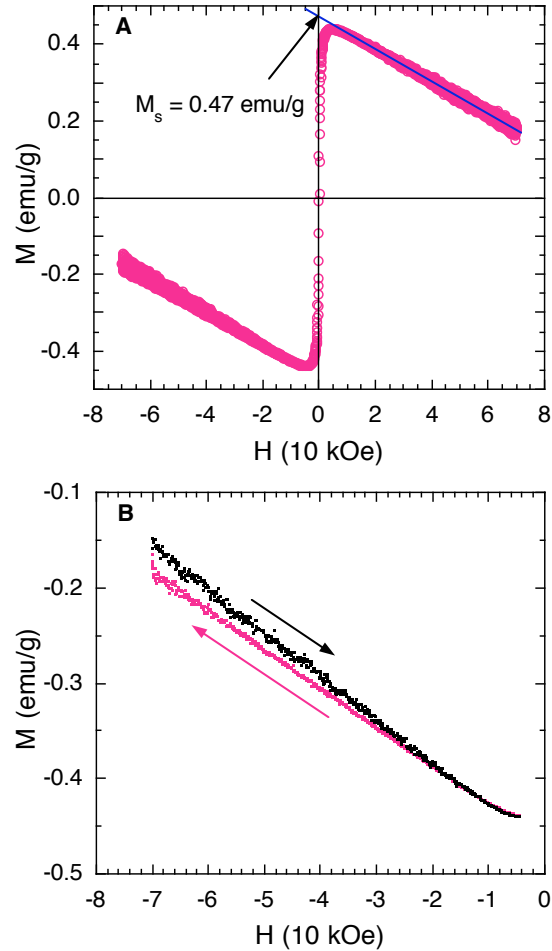

Fig. 3. Magnetic hysteresis loop of ferromagnetic nickel nanoparticles embedded in multi-walled carbon nanotubes. (**A**) Magnetization versus magnetic field at 320 K. The linear field dependence of the magnetization with a negative slope at $H > 10$ kOe is due to the diamagnetic contribution. The linear extrapolation to $H = 0$ yields $M_s = 0.47$ emu/g, which is a factor of 3.4 larger than the expected value (0.14 emu/g) from the measured Ni concentration and particle size. (**B**) An expanded view of the magnetization versus magnetic field. The magnetization is not reversible at a field higher than 20 kOe,

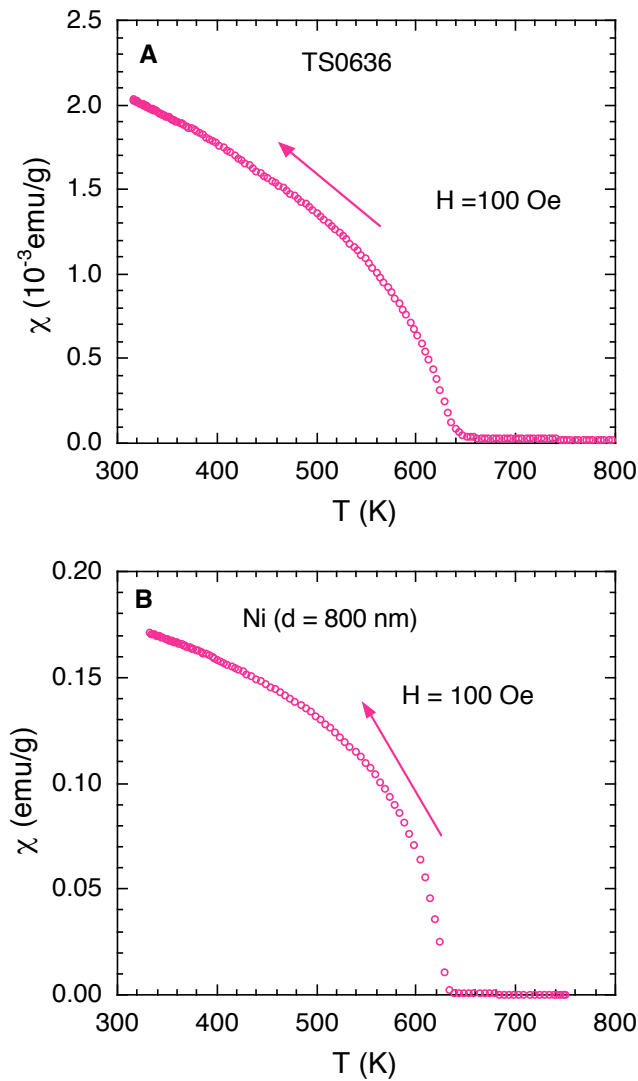

Fig. 4. Determination of the ferromagnetic transition temperature. (**A**) Field-cooled magnetic susceptibility of the MWCNT mat sample in a field of 100 Oe. There is only one ferromagnetic transition at about 630 K. (**B**) The field-cooled susceptibility for pure nickel particles ($d$ = 800 nm). Within the uncertainty in the sample temperature, the Curie temperature of the pure nickel particles is the same as that of the ferromagnetic nickel nanoparticles embedded in MWCNTs.